# Analysis of a Reduced-Communication Diffusion LMS Algorithm

Reza Arablouei [a] (corresponding author), Stefan Werner [b], Kutluyıl Doğançay [a], and Yih-Fang Huang [c]

[a] School of Engineering, University of South Australia, Mawson Lakes SA 5095, Australia
[b] Department of Signal Processing and Acoustics, School of Electrical Engineering, Aalto University, Espoo, Finland
[c] Department of Electrical Engineering, University of Notre Dame, Notre Dame, IN 46556, USA
E-mail: arary003@mymail.unisa.edu.au

**Abstract**: In diffusion-based algorithms for adaptive distributed estimation, each node of an adaptive network estimates a target parameter vector by creating an intermediate estimate and then combining the intermediate estimates available within its closed neighborhood. We analyze the performance of a reduced-communication diffusion least mean-square (RC-DLMS) algorithm, which allows each node to receive the intermediate estimates of only a subset of its neighbors at each iteration. This algorithm eases the usage of network communication resources and delivers a trade-off between estimation performance and communication cost. We show analytically that the RC-DLMS algorithm is stable and convergent in both mean and mean-square senses. We also calculate its theoretical steady-state mean-square deviation. Simulation results demonstrate a good match between theory and experiment.

## 1. Introduction

The diffusion strategies are effective methods for performing distributed estimation over adaptive networks. In a typical diffusion-based adaptive estimation algorithm, all network nodes concurrently generate individual intermediate estimates of a common target parameter vector using the data locally accessible to them. Then, the nodes communicate with all their immediate neighbors to exchange their intermediate estimates. Subsequently, each node fuses the intermediate estimates received from its neighborhood together with its own to create a new estimate. The procedure is repeated in all iterations [1]. This in-network cooperative processing helps the information propagate across the network so that all nodes can benefit from the observable data of the entire network. As a result, not only is the estimation performance of each network node significantly improved compared with when the nodes operate in isolation, but every node can asymptotically perform as well as in the often-hypothetical fully-connected case [2].

However, the benefits of diffusion-based algorithms come at the expense of increased internode communications. As all nodes transmit to and receive data from all their direct neighbors, depending on the connectivity density of the network, the total amount of required internode communications can become prohibitive. A large communication load may strain the valuable and usually-limited power, bandwidth, or hardware resources, particularly in wireless sensor networks. Moreover, in order to implement the conventional diffusion-based algorithms, each node should be able to communicate with all its neighbors simultaneously or within a certain time-frame. As a consequence, since different nodes can have different numbers of neighbors, they may require disparate hardware or consume power dissimilarly. This may in turn compromise the flexibility and efficiency of the network for ad-hoc deployment. Therefore, it is of practical importance to reduce the amount of internode communications in diffusion strategies while maintaining the benefits of cooperation.

There have been several attempts to reduce the communication complexity of the diffusion-based algorithms, particularly that of the diffusion least mean-square (DLMS) algorithm presented in [3]. In the probabilistic DLMS (P-DLMS) algorithm, each communication link is intermittently activated with a given probability [4]. Hence, the average amount of total internode communications taken place in the network is reduced. However, the total communication cost of the P-DLMS algorithm can vary in time. A performance analysis is presented in [4] that covers the P-DLMS algorithm as a special case. Nonetheless, it requires the calculation of two mean topology matrices by weighted-averaging over all the possible states of the network, which is unfeasible for not-so-small networks. In [5], an approach for dynamically optimizing the link probabilities of the P-DLMS algorithm is proposed. The single-link DLMS algorithms of [6] and [7] disconnect all links but one for each node at every iteration to reduce the communication overhead. Each one of these algorithms uses a different technique to select the neighbor with which each node communicates at any iteration by minimizing the steady-state network mean-square deviation (MSD). In [8], a set-theoretic diffusion-based algorithm is proposed, which can trade estimation performance and computational complexity for communication cost. The diffusion-based adaptive algorithm of [9] mitigates the communication load by exchanging either a scalar or a single information bit generated from random projections of the intermediate estimate vector of each node. The works of [10] and [11] utilize the concept of set-membership filtering [12] to alleviate the communication cost. In [13]-[15], two low-communication algorithms for adaptive distributed estimation are proposed that employ the notion of partial diffusion where each node transmits a part of the entries of its intermediate estimate vector to its neighbors at each iteration.

In this paper, we study the performance of a reduced-communication DLMS (RC-DLMS) algorithm for distributed estimation over adaptive networks. In the considered RC-DLMS algorithm, at each iteration, every node consults only a subset of its neighbors, i.e., receives the intermediate estimates of only a subset of its neighbors. This algorithm reduces the total amount of internode communications in the network relative to the DLMS algorithm, where the nodes always receive the intermediate estimates of all their neighbors, with limited degradation in performance. We analyze the performance of the RC-DLMS algorithm utilizing the energy conservation argument [16]. We establish its stability and convergence in the mean and mean-square senses. We also derive a theoretical expression for the steady-state MSD of the RC-DLMS algorithm and verify its accuracy via numerical simulations.

**2. Algorithm description**

*2.1. Diffusion least mean-square algorithm*

Consider a connected network of $K \in \mathbb{N}$ nodes that collectively aim to estimate a parameter vector denoted by $\mathbf{h} \in \mathbb{R}^{L \times 1}$ in an adaptive and collaborative manner. At each time instant $n \in \mathbb{N}$, every node $k \in \{1,2,\dots,K\}$ observes a regressor vector $\mathbf{x}_{k,n} \in \mathbb{R}^{L \times 1}$ and a scalar $y_{k,n} \in \mathbb{R}$ that are linearly related via

$$y_{k,n} = \mathbf{x}_{k,n}^T \mathbf{h} + v_{k,n} \tag{1}$$

where $v_{k,n} \in \mathbb{R}$ is the noise.

In the adapt-then-combine DLMS algorithm [3], each node produces an intermediate estimate using its previous estimate and most recent observed data (adaptation phase):

$$\mathbf{z}_{k,n} = \mathbf{w}_{k,n-1} + \mu_k \mathbf{x}_{k,n}\big(y_{k,n} - \mathbf{x}_{k,n}^T \mathbf{w}_{k,n-1}\big) \tag{2}$$

where $\mu_k \in \mathbb{R}_{>0}$ is the step-size at node $k$. After sharing its intermediate estimate with its neighbors, each node creates a new estimate by combining the intermediate estimates available within its neighborhood (consultation phase):

$$\mathbf{w}_{k,n} = c_{k,k} \mathbf{z}_{k,n} + \sum_{l \in \mathcal{N}_k} c_{k,l} \mathbf{z}_{l,n}. \tag{3}$$

The set $\mathcal{N}_k$ denotes the *open* neighborhood of node $k$, i.e., it comprises all nodes that are connected to node $k$ within one hop and excludes the node $k$ itself. The combination weights $\{c_{k,l} \in \mathbb{R}_{\geq 0}\}$ satisfy [1]

$$\forall k: \sum_{l=1}^{K} c_{k,l} = 1, \ c_{k,l} = 0 \text{ if } l \notin \mathcal{N}_k \cup \{k\}.$$

*2.2. Reduced-communication diffusion least mean-square algorithm*

At each iteration of the DLMS algorithm, every node receives the intermediate estimates of all its neighbors, which are $d_k = |\mathcal{N}_k|$ nodes where $|\cdot|$ is the cardinality operator and $d_k$ is called the degree of node $k$. To reduce the internode communications, one may allow each node to receive the intermediate estimates from $0 < m_k \leq d_k$ of its neighbors at each iteration. To realize this, we define a selection variable as $a_{k,l,n}$ that specifies the status of neighbor $l$ of node $k$ at time instant $n$. This variable can be either 0 or 1. Having $a_{k,l,n} = 1$ means that, at iteration $n$, node $k$ communicates with its neighboring node $l$ and receives its intermediate estimate to use in the consultation phase. On the other hand, $a_{k,l,n} = 0$ means that, at iteration $n$, node $k$ does not receive the intermediate estimate of its neighbor $l$.

We assume that the consulted neighbors of each node at each iteration are selected arbitrarily with equal probability. For node $k$, this probability is expressed as

$$p_k = E[a_{k,l,n}] = \frac{m_k}{d_k}.$$

Thus, we make the following remark regarding the considered reduced-communication scheme:

*R1*: At any node $k$ and iteration $n$, the neighbor-selection variable $a_{k,l,n}$ is statistically independent of the observed data $\mathbf{x}_{k,n}$ and $y_{k,n}$ as well as the noise $v_{k,n}$. Moreover, the selection probability $p_k$ is time-invariant and identical for all neighbors of node $k$.

When the intermediate estimates of only $m_k$ neighbors are received at node $k$, we may replace the unavailable intermediate estimates with the node's own intermediate estimate and change (3) to

$$\mathbf{w}_{k,n} = c_{k,k}\mathbf{z}_{k,n} + \sum_{l \in \mathcal{N}_k} c_{k,l}[a_{k,l,n}\mathbf{z}_{l,n} + (1 - a_{k,l,n})\mathbf{z}_{k,n}]. \tag{4}$$

Consequently, the considered reduced-communication DLMS (RC-DLMS) algorithm utilizes (2) in the adaptation phase and (4) in the consultation phase.

Note that the expressions (3) and (4) require an identical number of arithmetic operations, i.e., $(d_k + 1)L$ multiplications and $d_k L$ additions per iteration at any node $k$. Therefore, the RC-DLMS algorithm has the same computational complexity as the DLMS algorithm. Moreover, the overall communication cost of the RC-DLMS algorithm is constant over time since each node consults a fixed number of its neighbors at each iteration. On the other hand, the total communication cost of the P-DLMS algorithm can fluctuate in time due to the stochastic nature of the activation status of every link in this algorithm.

## 3. Performance analysis

We study the performance of the RC-DLMS algorithm in this section. The analysis covers the non-cooperative LMS and DLMS algorithms as the special cases of $m_k = 0$ and $m_k = d_k$, respectively.

*3.1. Assumptions*

For the analysis, we adopt the following assumptions, which are commonly used to facilitate the analytical studies [16]:

*A1*: The regressor vector $\mathbf{x}_{k,n}$ is temporally and spatially independent and

$$E[\mathbf{x}_{k,n}\mathbf{x}_{k,n}^T] = \mathbf{R}_k \in \mathbb{R}^{L \times L} \quad \forall k, n.$$

*A2*: The noise $v_{k,n}$ is independent of $\mathbf{x}_{k,n}$. In addition, it is temporally and spatially independent and

$$E[v_{k,n}] = 0 \text{ and } E[v_{k,n}^2] = \zeta_k^2 \in \mathbb{R} \quad \forall k, n.$$

*3.2. Network update equation*

Define

$$\check{\mathbf{z}}_{k,n} = \mathbf{z}_{k,n} - \mathbf{h},$$

$$\check{\mathbf{w}}_{k,n} = \mathbf{w}_{k,n} - \mathbf{h},$$

$$\check{\mathbf{w}}_n = [\check{\mathbf{w}}_{1,n}^T, \dots, \check{\mathbf{w}}_{K,n}^T]^T.$$

Subtracting $\mathbf{h}$ from both sides of (2) and (4) while using (1) gives

$$\check{\mathbf{z}}_{k,n} = (\mathbf{I}_L - \mu_k \mathbf{x}_{k,n}\mathbf{x}_{k,n}^T)\check{\mathbf{w}}_{k,n-1} + \mu_k \mathbf{x}_{k,n} v_{k,n},$$

$$\check{\mathbf{w}}_{k,n} = \left(1 - \sum_{l \in \mathcal{N}_k} a_{k,l,n} c_{k,l}\right)\check{\mathbf{z}}_{k,n} + \sum_{l \in \mathcal{N}_k} a_{k,l,n} c_{k,l} \check{\mathbf{z}}_{l,n},$$

which lead to

$$\check{\mathbf{w}}_n = \acute{\mathbf{B}}_n(\mathbf{I}_{LK} - \mathbf{M}\mathbf{X}_n)\check{\mathbf{w}}_{n-1} + \acute{\mathbf{B}}_n \mathbf{M} \mathbf{g}_n \tag{5}$$

where

$$\mathbf{M} = \text{blockdiag}\{\mu_1 \mathbf{I}_L, \cdots, \mu_K \mathbf{I}_L\},$$

$$\mathbf{X}_n = \text{blockdiag}\{\mathbf{x}_{1,n}\mathbf{x}_{1,n}^T, \cdots, \mathbf{x}_{K,n}\mathbf{x}_{K,n}^T\},$$

$$\mathbf{g}_n = [\mathbf{x}_{1,n}^T v_{1,n}, \dots, \mathbf{x}_{K,n}^T v_{K,n}]^T,$$

$$\acute{\mathbf{B}}_n = \mathbf{B}_n \otimes \mathbf{I}_L,$$

$$\mathbf{B}_n = \begin{bmatrix} b_{1,1,n} & \cdots & b_{1,K,n} \\ \vdots & \ddots & \vdots \\ b_{K,1,n} & \cdots & b_{K,K,n} \end{bmatrix},$$

$$b_{i,j,n} = \begin{cases} 1 - \sum_{l \in \mathcal{N}_i} a_{i,l,n} c_{i,l} & \text{if } j = i \\ a_{i,j,n} c_{i,j} & \text{if } j \in \mathcal{N}_i \\ 0 & \text{otherwise,} \end{cases}$$

and $\otimes$ denotes the Kronecker product.

*3.3. Mean stability*

From *R1* and *A1-A2*, we deduce the following corollary:

*C1*: The vector $\widecheck{\mathbf{w}}_{n-1}$ is statistically independent of $\acute{\mathbf{B}}_n$, $\mathbf{X}_n$, and $\mathbf{g}_n$. Moreover, $\acute{\mathbf{B}}_n$ is statistically independent of $\mathbf{X}_n$ and $\mathbf{g}_n$.

Taking the expectation on both sides of (5) while considering *C1* and *A2* results in

$$E[\widecheck{\mathbf{w}}_n] = (\bar{\mathbf{B}} \otimes \mathbf{I}_L)(\mathbf{I}_{LK} - \mathbf{MR})E[\widecheck{\mathbf{w}}_{n-1}]$$

where

$$\mathbf{R} = \text{blockdiag}\{\mathbf{R}_1, \cdots, \mathbf{R}_K\},$$

$$\bar{\mathbf{B}} = E[\mathbf{B}_n] = \begin{bmatrix} \bar{b}_{1,1,n} & \cdots & \bar{b}_{1,K,n} \\ \vdots & \ddots & \vdots \\ \bar{b}_{K,1,n} & \cdots & \bar{b}_{K,K,n} \end{bmatrix},$$

$$\bar{b}_{i,j,n} = \begin{cases} 1 - \sum_{l \in \mathcal{N}_i} p_i c_{i,l} & j = i \\ p_i c_{i,j} & j \in \mathcal{N}_i \\ 0 & \text{otherwise} \end{cases} = \begin{cases} 1 - p_i + p_i c_{i,i} & j = i \\ p_i c_{i,j} & j \in \mathcal{N}_i \\ 0 & \text{otherwise.} \end{cases}$$

All entries of $\bar{\mathbf{B}} \otimes \mathbf{I}_L$ are real non-negative and all its rows add up to unity as we have

$$(\bar{\mathbf{B}} \otimes \mathbf{I}_L)\mathbf{1}_{LK} = E[\mathbf{B}_n \mathbf{1}_K] \otimes \mathbf{1}_L = \mathbf{1}_K \otimes \mathbf{1}_L = \mathbf{1}_{LK}.$$

Therefore, $\bar{\mathbf{B}} \otimes \mathbf{I}_L$ is right-stochastic. As a result, similar to the DLMS algorithm, the mean stability and asymptotic unbiasedness of the RC-DLMS algorithm is ensured if the spectral radius of $\mathbf{I}_{LK} - \mathbf{MR}$ is smaller than one [3]. This can be realized by choosing the step-size of each node $k$ such that

$$0 < \mu_k < \frac{2}{\lambda_{\max}\{\mathbf{R}_k\}} \tag{6}$$

where $\lambda_{\max}\{\mathbf{R}_k\}$ is the largest eigenvalue of $\mathbf{R}_k$.

*3.4. Variance relation*

Denote an arbitrary symmetric nonnegative-definite matrix by **S** and define the squared-weighted Euclidean norm of a vector $\mathfrak{b}$ with a weighting matrix $\mathfrak{U}$ as

$$\|\mathfrak{b}\|_{\mathfrak{U}}^2 = \|\mathfrak{b}\|_{\text{vec}\{\mathfrak{U}\}}^2 = \mathfrak{b}^T \mathfrak{U} \mathfrak{b}.$$

Taking the expectation of the squared-weighted Euclidean norm on both sides of (5) with *C1* in mind gives the following weighted variance relation:

$$E[\|\widetilde{\mathbf{w}}_n\|_{\mathbf{S}}^2] = E[\|\widetilde{\mathbf{w}}_{n-1}\|_{\mathbf{T}}^2] + E[\mathbf{g}_n^T \mathbf{M} \acute{\mathbf{B}}_n^T \mathbf{S} \acute{\mathbf{B}}_n \mathbf{M} \mathbf{g}_n] \tag{7}$$

$$\mathbf{T} = E[(\mathbf{I}_{LK} - \mathbf{M}\mathbf{X}_n) \acute{\mathbf{B}}_n^T \mathbf{S} \acute{\mathbf{B}}_n (\mathbf{I}_{LK} - \mathbf{M}\mathbf{X}_n)]. \tag{8}$$

Applying the vectorization operator to (8) together with using the property [17]

$$\text{vec}\{\mathfrak{U}\mathfrak{B}\mathfrak{C}\} = (\mathfrak{C}^T \otimes \mathfrak{U})\text{vec}\{\mathfrak{B}\}$$

yields

$$\text{vec}\{\mathbf{T}\} = \mathbf{F}\mathbf{D}\mathbf{s} \tag{9}$$

where

$$\mathbf{F} = E[(\mathbf{I}_{LK} - \mathbf{M}\mathbf{X}_n) \otimes (\mathbf{I}_{LK} - \mathbf{M}\mathbf{X}_n)]$$
$$\approx (\mathbf{I}_{LK} - \mathbf{M}\mathbf{R}) \otimes (\mathbf{I}_{LK} - \mathbf{M}\mathbf{R}), \tag{10}$$

$$\mathbf{D} = E[\acute{\mathbf{B}}_n^T \otimes \acute{\mathbf{B}}_n^T],$$

$$\mathbf{s} = \text{vec}\{\mathbf{S}\}.$$

The approximation in the second line of (10) is reasonable when the step-sizes are sufficiently small.

The transpose of **D** is calculated as

$$\mathbf{D}^T = E[\mathbf{B}_n \otimes \mathbf{I}_L \otimes \mathbf{B}_n \otimes \mathbf{I}_L] = \begin{bmatrix} \mathbf{I}_L \otimes E[b_{1,1,n}\mathbf{B}_n] & \cdots & \mathbf{I}_L \otimes E[b_{1,K,n}\mathbf{B}_n] \\ \vdots & \ddots & \vdots \\ \mathbf{I}_L \otimes E[b_{K,1,n}\mathbf{B}_n] & \cdots & \mathbf{I}_L \otimes E[b_{K,K,n}\mathbf{B}_n] \end{bmatrix} \otimes \mathbf{I}_L$$

using

$$E[b_{i,j,n}b_{k,l,n}] = \begin{cases} 1 - p_i(1 - c_{i,i}) - p_k(1 - c_{k,k}) + \sum_{r \in \mathcal{N}_i} \sum_{u \in \mathcal{N}_k} c_{i,r}c_{k,u}E[a_{i,r,n}a_{k,u,n}] & i = j\ \&\ k = l \\ p_k c_{k,l} - c_{k,l} \sum_{r \in \mathcal{N}_i} c_{i,r}E[a_{i,r,n}a_{k,l,n}] & i = j\ \&\ l \in \mathcal{N}_k \\ p_i c_{i,j} - c_{i,j} \sum_{r \in \mathcal{N}_k} c_{k,r}E[a_{i,j,n}a_{k,r,n}] & j \in \mathcal{N}_i\ \&\ k = l \\ c_{i,j}c_{k,l}E[a_{i,j,n}a_{k,l,n}] & j \in \mathcal{N}_i\ \&\ l \in \mathcal{N}_k \\ 0 & \text{otherwise} \end{cases}$$

and

$$E[a_{i,j,n}a_{k,l,n}] = \begin{cases} p_i & i = k \,\&\, j = l \\ 0 & i = k \,\&\, j \neq l \,\&\, d_i = 1 \\ p_i \dfrac{m_i - 1}{d_i - 1} & i = k \,\&\, j \neq l \,\&\, d_i > 1 \\ p_i p_k & i \neq k. \end{cases}$$

In view of *R1*, the property [17]

$$\text{tr}\{\mathfrak{A}^T\mathfrak{B}\} = \text{vec}^T\{\mathfrak{B}\}\text{vec}\{\mathfrak{A}\},$$

and the fact that **S** is symmetric and deterministic, we have

$$E[\mathbf{g}_n^T \mathbf{M}\acute{\mathbf{B}}_n^T \mathbf{S}\acute{\mathbf{B}}_n \mathbf{M}\mathbf{g}_n] = \text{vec}^T\{\mathbf{H}\}\mathbf{Ds} \tag{11}$$

where

$$\mathbf{H} = \mathbf{M}E[\mathbf{g}_n\mathbf{g}_n^T]\mathbf{M} = \text{blockdiag}\{\mu_1^2\zeta_1^2\mathbf{R}_1, \cdots, \mu_K^2\zeta_K^2\mathbf{R}_K\}.$$

Substituting (9) and (11) into (7) gives

$$E[\|\widecheck{\mathbf{w}}_n\|_\mathbf{s}^2] = E[\|\widecheck{\mathbf{w}}_{n-1}\|_{\mathbf{FDs}}^2] + \text{vec}^T\{\mathbf{H}\}\mathbf{Ds}. \tag{12}$$

*3.5. Mean-square stability*

The recursion (12) is stable if the matrix **FD** is stable [16]. The entries of **D** are all real-valued and non-negative. In addition, we have

$$\mathbf{D}^T\mathbf{1}_{L^2K^2} = E[(\mathbf{B}_n \otimes \mathbf{I}_L)\mathbf{1}_{LK} \otimes (\mathbf{B}_n \otimes \mathbf{I}_L)\mathbf{1}_{LK}] = E[\mathbf{B}_n\mathbf{1}_K \otimes \mathbf{1}_L \otimes \mathbf{B}_n\mathbf{1}_K \otimes \mathbf{1}_L] = \mathbf{1}_{L^2K^2}.$$

This means that **D** is left-stochastic. Therefore, the RC-DLMS algorithm is stable in the mean-square sense and converges to a steady state if **F** is stable. Considering the approximation in (10), **F** is stable and consequently the RC-DLMS algorithm is mean-square stable when the step-sizes satisfy (6).

*3.6. Steady-state mean-square deviation*

Using (12), at the steady state, we can write

$$\lim_{n\to\infty} E\left[\|\widecheck{\mathbf{w}}_n\|_{(\mathbf{I}_{L^2K^2}-\mathbf{FD})\mathbf{s}}^2\right] = \text{vec}^T\{\mathbf{H}\}\mathbf{Ds}. \tag{13}$$

Setting

$$\mathbf{s} = (\mathbf{I}_{L^2K^2} - \mathbf{FD})^{-1}\text{vec}\{\mathbf{J}_K^{k,k} \otimes \mathbf{I}_L\}$$

in (13), the steady-state MSD of node $k$, defined by

$$\eta_k = \lim_{n\to\infty} E\left[\|\widecheck{\mathbf{w}}_{k,n}\|^2\right],$$

can be calculated as

$$\eta_k = \text{vec}^T\{\mathbf{H}\}\mathbf{D}(\mathbf{I}_{L^2K^2} - \mathbf{FD})^{-1}\text{vec}\{\mathbf{J}_K^{k,k} \otimes \mathbf{I}_L\}.$$

All the entries of $\mathbf{J}_K^{k,k} \in \mathbb{R}^{K \times K}$ are zero except the $(k,k)$th entry that is one. Similarly, the steady-state network MSD is defined and calculated as

$$\eta = \frac{1}{K}\sum_{k=1}^{K} \eta_k = \frac{1}{K}\text{vec}^T\{\mathbf{H}\}\mathbf{D}(\mathbf{I}_{L^2K^2} - \mathbf{FD})^{-1}\text{vec}\{\mathbf{I}_{LK}\}.$$

Note that the stability of $\mathbf{FD}$ implies that $\mathbf{I}_{L^2K^2} - \mathbf{FD}$ is invertible.

The analysis presented in this section can also apply to the P-DLMS algorithm if we consider $E[a_{k,l,n}] = p_{k,l}$ and $E[a_{i,j,n}a_{k,l,n}] = \begin{cases} p_{i,j} & i = k \ \& \ j = l \\ p_{i,j}p_{k,l} & \text{otherwise} \end{cases}$ where $p_{k,l}$ is the probability of activeness of the link from node $l$ to node $k$.

## 4. Simulations

We consider an adaptive network of $K = 20$ nodes that are arbitrarily connected to each other and each node has between one to seven neighbors excluding itself. On average, each node is connected to four other nodes. The nodes collectively identify a parameter vector of length $L = 4$. The regressor at each node is zero-mean multivariate Gaussian with an arbitrary covariance matrix. The additive noise at each node is also zero-mean Gaussian. The regressors and the noise are temporally and spatially independent of each other. In Fig. 1, we show the trace of the regressor covariance matrix and the variance of the noise at each node. We obtain the experimental results by taking the ensemble average over $10^4$ independent trials and the steady-state quantities by averaging over 500 steady-state values. We also use the same step-size, denoted by $\mu$, in all nodes. In the RC-DLMS algorithm, we determine the number of neighbors with which each node communicates at each iteration to receive their intermediate estimates via $m_k = \min(M, d_k)$ where $0 \leq M \in \mathbb{N} \leq d_k$ specifies the maximum number of consulted neighbors of every node at each iteration.

In Fig. 2, we plot the time-evolution of the network MSD of the RC-DLMS algorithm for different values of $M$ when $\mu = 0.01$. We use the relative-degree weights [1] for $\{c_{k,l}\}$ in the consultation phase. Note that the RC-DLMS algorithm becomes the non-cooperative LMS algorithm when $M = 0$ and the DLMS algorithm when $M = 7$. In Fig. 3, we compare the theoretical and experimental values of the steady-state network MSD of the RC-DLMS algorithm for different values of $M$ and $\mu$. In Fig. 4, we compare the theoretical and experimental values of the steady-state MSD of all nodes for different values of $M$ when $\mu = 0.01$. We observe in Figs. 2-4 that the RC-DLMS algorithm provides an effective trade-off between performance and communication cost. Moreover, there is a good agreement between the theoretical and experimental steady-state MSD values for a wide range of $\mu$ and $M$.

## 5. Conclusion

We examined a reduced-communication diffusion least mean-square (RC-DLMS) algorithm, which enables reduced internode communications by allowing each node to receive the intermediate estimates of a subset of its neighbors at every iteration. We studied the convergence performance of this algorithm and predicted its steady-state mean-square deviation. Simulations results confirmed the accuracy of the theoretical predictions. The presented analysis provided valuable insights into the performance of the RC-DLMS algorithm and demonstrated that it offers an effective trade-off between estimation performance and communication cost.

## References

[1] A. H. Sayed, "Diffusion adaptation over networks," in *Academic Press Library in Signal Processing*, vol. 3, R. Chellapa and S. Theodoridis, Eds., Academic Press, 2013, pp. 323-454.

[2] A. H. Sayed, "Adaptation, learning, and optimization over networks," *Foundations and Trends in Machine Learning*, vol. 7, no. 4-5, pp. 311–801, Jul. 2014.

[3] F. S. Cattivelli and A. H. Sayed, "Diffusion LMS strategies for distributed estimation," *IEEE Trans. Signal Process.*, vol. 58, pp. 1035–1048, Mar. 2010.

[4] C. G. Lopes and A. H. Sayed, "Diffusion adaptive networks with changing topologies," in *Proc. Int. Conf. Acoust. Speech Signal Process.*, Las Vegas, USA, Apr. 2008, pp. 3285–3288.

[5] N. Takahashi and I. Yamada, "Link probability control for probabilistic diffusion least-mean squares over resource-constrained networks," in *Proc. Int. Conf. Acoust. Speech Signal Process.*, Dallas, USA, Mar. 2010, pp. 3518–3521.

[6] X. Zhao and A. H. Sayed, "Single-link diffusion strategies over adaptive networks," in *Proc. Int. Conf. Acoust. Speech Signal Process.*, Kyoto Japan, Mar. 2012, pp. 3749–3752.

[7] Ø. L. Rørtveit, J. H. Husøy, and A. H. Sayed, "Diffusion LMS with communication constraints," in *Proc. IEEE Asilomar Conf. Signals Syst. Comput.*, Pacific Grove, USA, Nov. 2010, pp. 1645–1649.

[8] S. Chouvardas, K. Slavakis, and S. Theodoridis, "Trading off complexity with communication costs in distributed adaptive learning via Krylov subspaces for dimensionality reduction," *IEEE J. Sel. Topics Signal Process.*, vol. 7, pp. 257-273, Apr. 2013.

[9] M. O. Sayin and S. S. Kozat, "Compressive diffusion strategies over distributed networks for reduced communication load," *IEEE Trans. Signal Process.*, vol. 62, no. 20, pp. 5308–5323, Oct. 15, 2014.

[10] S. Werner, Y.-F. Huang, M. L. R. de Campos, and V. Koivunen, "Distributed parameter estimation with selective cooperation," in *Proc. Int. Conf. Acoust. Speech Signal Process.*, Taipei, Taiwan, Apr. 2009, pp. 2849-2852.

[11] S. Werner, M. Mohammed, Y.-F. Huang, and V. Koivunen, "Decentralized set-membership adaptive estimation for clustered sensor networks," in *Proc. Int. Conf. Acoust. Speech Signal Process.*, Las Vegas, USA, Apr. 2008, pp. 3573–3576.

[12] S. Gollamudi, S. Nagaraj, S. Kapoor, and Y.-F. Huang, "Set-membership filtering and a set-membership normalized LMS algorithm with an adaptive step size," *IEEE Signal Process. Lett.*, vol. 5, pp. 111–114, May 1998.

[13] R. Arablouei, S. Werner, Y.-F. Huang, and K. Doğançay, "Distributed least mean-square estimation with partial diffusion," *IEEE Trans. Signal Process.*, vol. 62, no. 2, pp. 472-484, Jan. 2014.

[14] R. Arablouei, K. Doğançay, S. Werner, and Y.-F. Huang, "Adaptive distributed estimation based on recursive least-squares and partial diffusion," *IEEE Trans. Signal Process.*, vol. 62, no. 14, pp. 3510-3522, Jul. 2014.

[15] R. Arablouei, S. Werner, and K. Doğançay, "Partial-diffusion recursive least-squares estimation over adaptive networks," in *Proc. IEEE Int. Workshop Computational Advances Multi-Sensor Adaptive Process.*, Saint Martin, Dec. 2013, pp. 89-92.

[16] A. H. Sayed, *Adaptive Filters*, Hoboken, NJ: Wiley, 2008.

[17] K. M. Abadir and J. R. Magnus, *Matrix Algebra*, NY: Cambridge Univ. Press, 2005.

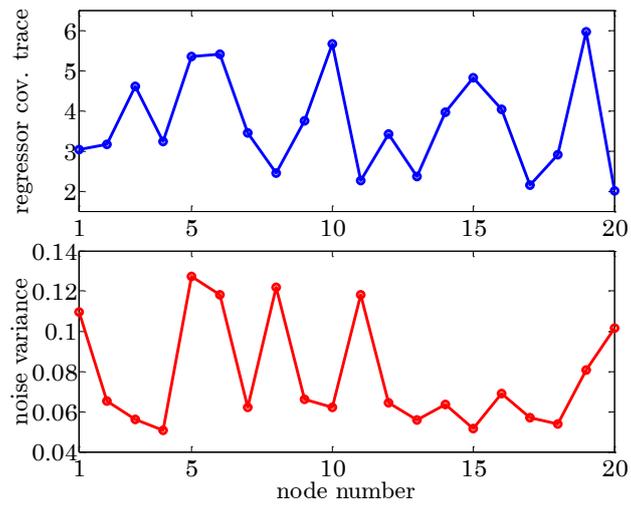

Fig. 1. Trace of the regressor covariance matrix and variance of the noise at each node.

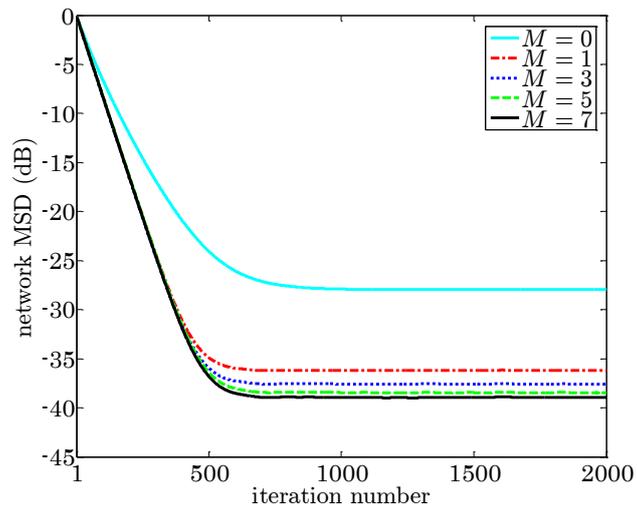

Fig. 2. Network MSD curves of the RC-DLMS algorithm with different values of $M$ when $\mu = 0.01$.

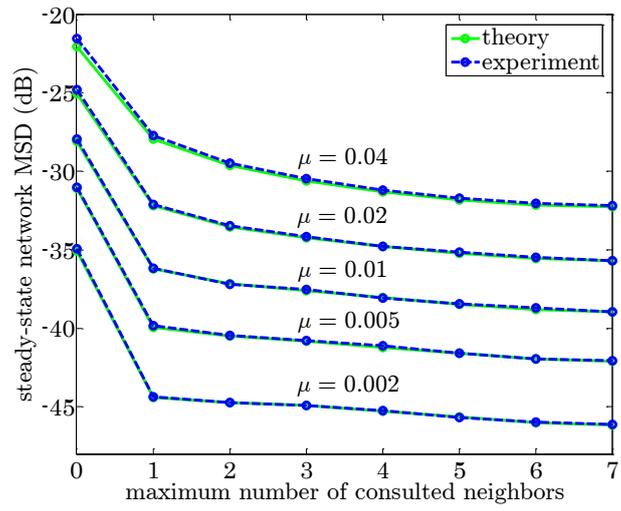

Fig. 3. Theoretical and experimental steady-state network MSDs of the RC-DLMS algorithm versus $M$ for different values of $\mu$.

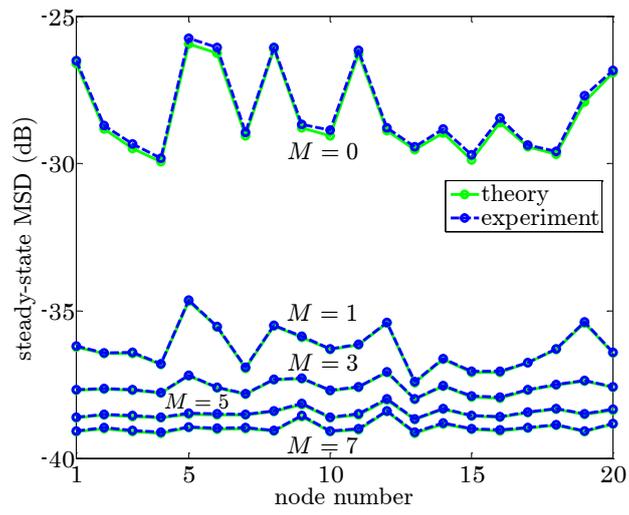

Fig. 4. Theoretical and experimental steady-state MSDs of the RC-DLMS algorithm at each node for different values of $M$ when $\mu = 0.01$.